\documentclass[12pt]{article}

\newlength{\dinwidth}
\newlength{\dinmargin}
\DeclareMathAlphabet{\mathsfsl}{OT1}{cmss}{m}{sl}
\DeclareMathAlphabet{\mathscr}{U}{eus}{m}{n}
\DeclareMathAlphabet{\matheur}{U}{eur}{m}{n}
\SetMathAlphabet{\matheur}{bold}{U}{eur}{b}{n}
\newcommand{\nn}{\nonumber \\}
\newcommand{\beq}{\begin{equation}} 
\newcommand{\eeq}{\end{equation}}
\newcommand{\bea}{\begin{eqnarray}}
\newcommand{\eea}{\end{eqnarray}}

\newcommand{\bruch}[2]{{\textstyle \frac{#1}{#2}}}
\newcommand{\A}{\alpha}
\newcommand{\B}{\beta}

\newcommand{\D}{\delta}

\newcommand{\G}{\Gamma}

\renewcommand{\a}{\alpha}
\renewcommand{\b}{\beta}

\renewcommand{\d}{\delta}

\newcommand{\eqn}[1]{(\ref{#1})}
\newcommand{\ft}[2]{{\textstyle\frac{#1}{#2}}}
\newcommand{\e}{\epsilon}

\newcommand{\m}{\mu}

\newcommand{\n}{\nu}

\newcommand{\AC}{{\mathcal{A}}}
\newcommand{\BC}{{\mathcal{B}}}
\newcommand{\CC}{{\mathcal{C}}}
\newcommand{\DC}{{\mathcal{D}}}

\newcommand{\VC}{{\mathcal{V}}}

\newcommand{\PC}{{\mathcal{P}}}
\newcommand{\QC}{{\mathcal{Q}}}

\newcommand{\s}{{\rm SO(16)}}
\newcommand{\so}{{\rm SO}(1,2) \times {\rm SO}(16)}
\newcommand{\sa}{{\rm SO}(1,2) \times {\rm SO}(8)}
\newcommand{\SB}{{\rm SO}(1,3) \times {\rm SO}(7)}

\newcommand{\su}{{\rm SO}(1,3) \times {\rm SU}(8)}

\newcommand{\AmS}{{\protect\the\textfont2
  A\kern-.1667em\lower.5ex\hbox{M}\kern-.125emS}}

\begin{document}
\thispagestyle{empty}
\renewcommand{\thefootnote}{\fnsymbol{footnote}}
\begin{flushright} hep-th/0011239\\[1mm]
   AEI--2000--072\\
   ITP-UU-00/29   \\
   SPIN-00/27  
\end{flushright}
\vspace*{1.0cm}
\begin{center}
 {\LARGE \bf Hidden Symmetries, Central Charges and All That\footnotemark
\footnotetext{Based on invited talks given at the G\"ursey Memorial
Conference II, Istanbul, June 2000} 
}\\
 \vspace*{1cm}
 {{\bf B. de Wit$^1$ and H. Nicolai$^2$} \\
 \vspace*{6mm}
 $^1$Institute for Theoretical Physics and Spinoza Institute\\
Utrecht University, Utrecht, The Netherlands\\[2mm]
     $^2$Max-Planck-Institut f\"ur Gravitationsphysik\\
     Albert-Einstein-Institut \\
     M\"uhlenberg 1, D-14476 Golm, Germany} \\
 \vspace*{1cm}
\begin{minipage}{11cm}\footnotesize
\textbf{Abstract:}
In this review we discuss hidden symmetries of
toroidal compactifications of eleven-dimensional supergravity. We
recall alternative versions of this theory which exhibit traces of
the hidden symmetries when still retaining the massive Kaluza-Klein
states. We reconsider them in the broader perspective of M-theory
which incorporates a more  extended variety of BPS states. We also
argue for a new  
geometry that may underly these theories. All our arguments point
towards an extension of the number of space-time coordinates beyond
eleven. 
\end{minipage}
\end{center}
\setcounter{footnote}{0}
\vspace*{0.2cm}
       
\section{Introduction}
One of the key problems in understanding superstring theory at 
the non-perturbative level is the question of its fundamental
underlying symmetry. Although we still do not know what this
symmetry is, it has become clear in recent years that $d=11$ 
supergravity \cite{CJS} will play a central role in this endeavor
\cite{HullTownsend,Townsend,Witten}. The hidden global
symmetries arising 
in the dimensional reduction of this theory to lower dimensions
\cite{CJ} have been conjectured to also appear in M-theory 
\cite{HullTownsend}, albeit in a discrete version, and only for
toroidal compactifications. Over the past few years it has also 
become clear that BPS states and supermultiplets are of essential
importance in this context.

In this contribution we review the status of ``hidden symmetries''
in supergravity and superstring theory, and their connection with
central charges from a point of view which is somewhat different 
from the one usually taken. Namely, we will base our considerations 
on some older work \cite{dewnic1,nic1} where it was shown that
traces of the hidden ${\rm E}_{n(n)}$ symmetries of dimensionally 
reduced supergravity \cite{CJ} remain in eleven dimensions. These
considerations will lead us to conjecture the existence of effective 
field theories also encompassing the (non-perturbative) BPS degrees 
of freedom that are fully compatible with the hidden symmetries (a
subset of which will be the Kaluza-Klein states). We will refer to
these theories as ``BPS-extended  
supergravities'' \cite{strings00}, as they would be of a new type. In
particular, they would live in a higher-dimensional space, such that
the central charges would be associated with certain extra dimensions,
in a way similar to the central charges that originate from the internal
momenta in a Kaluza-Klein compactification. We will also argue for
a ``hidden exceptional geometry'' underlying $d=11$ supergravity
and/or M-theory.

\section{Central Charges from Eleven Dimensions}
In $D=11$ space-time dimensions the anticommutator of the supercharges
decomposes as follows  
\beq
\{ Q_\A , \bar Q_\B \} = \G^M_{\A\B} P_M + \ft12 \G^{MN}_{\A\B} Z_{MN}
    + \ft1{5!} \G^{MNPQR}_{\A\B} Z_{MNPQR} \,. \label{D11-algebra}
\eeq
Here the $P_M$ denote the 11-dimensional momentum operators and $Z_{MN}$ and
$Z_{MNPQR}$ are the charges associated with two- and
five-branes.  The $P_M$, $Z_{MN}$ and $Z_{MNPQR}$ represent $11+55+462
= 528$ components and thus generally parametrize the anti-commutator
on the right-hand side. Upon 
compactification on a torus $T^n$, where 
$n={11-d}$,  we are dealing with an extended  supersymmetry algebra in
$d$ space-time dimensions. In this way one 
obtains corresponding centrally extended maximal supersymmetry
algebras in lower dimensions. A priori these central charges transform
according to representations of an internal ${\rm SO}(n)$, but in fact
there is a bigger group to which they can be assigned, namely to the 
automorphism group ${\rm H}_{\rm R}$ of the supersymmetry algebra that
acts on the supercharges and commutes with the $d$-dimensional Lorentz
group (for a classification, see e.g. \cite{deWitLouis}). It turns out,
however, that this assignment can be further 
extended (although not in general as we shall discuss below), namely to
representations of the hidden symmetry group G. This is shown in
table~1 (see \cite{ObersPioline} for a comprehensive review).

\begin{table}
\begin{center}
\begin{tabular}{l l l l}\hline
$d$ & ${\rm G}$ & ${\rm H}_{\rm R}$ & representations \\ \hline
 9   & ${\rm SL}(2,{\bf R}) \times {\rm SO}(1,1)$  & SO(2) &
$({\bf{2}},{\bf 1}) 
\oplus ({\bf{1}},{\bf{1}})$\\ 
 8   & ${\rm SL}(3,{\bf R})\times {\rm SL}(2,{\bf R})$   & U(2) &
$(\bf{3},\bf{2})$\\ 
 7    & ${\rm E}_{4(4)}\equiv {\rm SL}(5,{\bf R})$ & USp(4)     &
$\bf{10}$\\ 
 6    & ${\rm E}_{5(5)}\equiv {\rm SO}(5,5)$ &${\rm USp}(4)\times {\rm
USp}(4)$    & ${\bf{16}}\to ({\bf 4},{\bf 4})$\\
 5    & ${\rm E}_{6(6)}$ & USp(8)     & $\bf{27}\oplus \bf{1}$\\
 4    & ${\rm E}_{7(7)}$ &SU(8)   & ${\bf{56}}\to {\bf 28}\oplus
\overline{\bf 28}$\\\hline 
\end{tabular}
\end{center}
\caption{The hidden symmetry groups G and the groups ${\rm H}_{\rm R}$
for $4\leq d \leq 9$ with the G-representations of the pointlike
central charges. We indicate the branching into ${\rm H}_{\rm R}$
representations for $d=6$ and 4. }
\label{CC1}  
\end{table}

According to \eqn{D11-algebra} the pointlike central charges in $d$
space-time dimensions can be classified,
respectively,  into ``Kaluza-Klein central charges'' $P_m$ originating
from the $D=11$ momentum operator $P_M$,  
and ``winding central charges'' $Z_{mn}$ and $Z_{mnpqr}$ originating
from $Z_{MN}$ and $Z_{MNPQR}$.   In this way we get (the number of
central charge components is given in brackets)
\beq
\begin{array}{cclcl}
d=9 &\rightarrow & P_m \, [2]&  ,  & Z_{mn}\, [1]  \\
d=8 &\rightarrow & P_m \, [3]&,&  Z_{mn}\, [3]
\\ 
d=7 &\rightarrow & P_m \, [4]&,& Z_{mn} \, [6] \\
d=6 &\rightarrow & P_m \, [5]&,& Z_{mn}[10] \oplus
                          Z_{mnpqr} \, [1] \\
d=5 &\rightarrow & P_m \, [6] &,& Z_{mn} \, [15] \oplus
      Z_{mnpqr} [6] \oplus Z_{\mu\nu\rho\sigma\tau} [1] \\
d=4 &\rightarrow & P_m \, [7] &,& Z_{mn} [21] \oplus
      Z_{mnpqr} [21] \oplus Z_{m\mu\nu\rho\sigma} [7] \\
d=3 &\rightarrow & P_m \, [8] &,& Z_{mn}[28] \oplus
      Z_{mnpqr} [56] \oplus Z_{mn\mu\nu\rho} [28] \\
d=2 &\rightarrow & P_m \, [9] &,& Z_{mn}[36] \oplus Z_{\mu\nu} [1]
    \oplus  Z_{mnpqr} [126] \oplus Z_{mnp\mu\nu} [84] 
\end{array} 
\eeq
where we used space-time indices $\mu,\nu,\ldots=0,\ldots,d-1$ and
internal indices  $m,n,\ldots= 1,\ldots,11-d$. These central charges
transform according to representations of the group ${\rm H}_{\rm
R}$. In most cases this representation is irreducible (exceptions
occur for  $d=2,5$ and 9). 

As indicated above, for $d\geq 4$ the central charges combine into
representations of the bigger (hidden) symmetry groups ${\rm
E}_{n(n)}$. We already listed these representations in table~1.
Below  $d=4$ the pointlike central charges no longer fit into
representations of G, but only into
representations of the ${\rm H}_{\rm R}$: for $d=3$, we get the $\bf
120$ of SO(16) (rather than a representation of ${\rm E}_{8(8)}$), and
for $d=2$ 
we have $\bf{1} \oplus \bf{120}\oplus \bf{135}$ of ${\rm SO}(16)$
(rather than a representation of ${\rm E}_{9(9)}$). In the latter case, 
the centrally extended maximal superalgebra in $d=2$ is given by 
\beq
\{ Q^I_\pm , Q^J_\pm \} = \D^{IJ} P_\pm\,, \qquad
\{ Q^I_+ , Q^J_-\} = Z^{IJ}\,,
\eeq
in terms of (one-component) Majorana-Weyl spinors
with indices $I,J=1,\ldots,16$. While left- or right-moving
BPS states corresponding to elimination of  
either $Q_+^I$ or $Q_-^I$ (i.e. $(16,0)$ or $(0,16)$ supersymmetry)
are massless and do not involve central charges, massive states
involving the 256 central charges $Z^{IJ}$ have not been considered 
in the literature so far.

Similar considerations apply to the central charges with Lorentz
indices, such as the ``stringlike'' central charges
with one space-time index; these charges are carried by one-dimensional
extended objects (strings) rather than point particles. We get
\bea
\begin{array}{ccl}
d=9 &\rightarrow & Z_{\mu m} \, [2]   \\
d=8 &\rightarrow & Z_{\mu m} \, [3]   \\
d=7 &\rightarrow & Z_{\mu m} \, [4] \oplus Z_{\mu mnpq} \, [1] \\
d=6 &\rightarrow & Z_{\mu m}\, [5] \oplus
       Z_{\mu mnpq} \, [5] \oplus Z_{\mu\nu\rho\sigma\tau} \, [1] \\
d=5 &\rightarrow & Z_{\mu m} \, [6] \oplus
   Z_{\mu mnpq} \, [15] \oplus Z_{\mu\nu\rho\sigma m} [6] \\
d=4 &\rightarrow & Z_{\mu m} [7] \oplus  Z_{\mu mnpq} \, [35] 
\oplus Z_{\mu\nu\rho mn} [21] \\
d=3 &\rightarrow & Z_{\mu\nu}[1] \oplus Z_{\mu m}[8] \oplus
      Z_{\mu mnpq} \, [70] \oplus Z_{\mu\nu mnp} [56] \\
d=2 &\rightarrow & Z_{\mu m}[9] \oplus Z_{\mu mnpq} [126] 
\end{array}
\eea
Again these charges transform according to the group ${\rm H}_{\rm R}$. 
For $d\geq5$ they  also  fit into representations of G; these cases
are listed in table~2. For $d=4$ we have a $\bf 63$ of 
SU(8) (and not a representation of ${\rm E}_{7(7)}$). For $d=3,2$ we
have a $\bf 135$ representation of SO(16) (and not a representation 
of ${\rm E}_{8(8)}$ or ${\rm E}_{9(9)}$, respectively). The same 
pattern is seen for the two-brane charges, which only fit into
representations of the hidden symmetry group for $d\geq6$. For higher
brane charges the phenomenon does not occur, perhaps because the
symmetry groups become simpler and so do the representations. It is
noteworthy that the failure of the pointlike, stringlike and two-brane
charges to constitute representations of G takes place at those
dimensions where the corresponding gauge fields (with rank 1, 2, and 3,
respectively) can be dualized to scalars or where they cease to
exist altogether. 

\begin{table}
\begin{center}
\begin{tabular}{l l l}\hline
$d$ & ${\rm G}$ & representations \\\hline
 9   & ${\rm SL}(2,{\bf R})\times {\rm SO}(1,1)$        & $\bf{2}$\\
 8   & ${\rm SL}(3,{\bf R})\times {\rm SL}(2,{\bf R})$ &
$(\bf{3},\bf{1})$\\ 
 7    & ${\rm SL}(5,{\bf R})$      & $\bf{5}$\\
 6    & ${\rm SO}(5,5)$   & ${\bf{10}}\oplus {\bf{1}}\to 
({\bf 5},{\bf 1})\oplus ({\bf 1},{\bf 5}) \oplus ({\bf 1},{\bf 1})$\\ 
 5    & ${\rm E}_{6(6)}$      & $\overline{\bf{27}} $\\\hline
\end{tabular}
\end{center}
\caption{Stringlike central charges and their representations 
in dimensions $d\geq 5$. We indicate the branching into ${\rm H}_{\rm
R}$ representations for $d=6$. }\label{CC2} 
\end{table}

It is evident that the toroidal compactifications of maximal
supergravity are incomplete, even when retaining the Kaluza-Klein
states. While correctly describing the dynamics of 
the massless sectors they do not incorporate the full set of BPS
states, simply because the charges $Z_{MN}$ and $Z_{MNPQR}$ remain
zero. It is obvious that some of these missing charges can be generated by
solitonic solutions (or they can be introduced explicitly into the
field theory). Nevertheless there seem to remain certain 
deficiencies in the spectrum. For example, there is
a 3-brane central charge transforming according to $({\bf 10},{\bf 1})
\oplus ({\bf 1},{\bf 10})$ of ${\rm USp}(4)\times {\rm USp}(4)$ in
$d=6$. However, there are no massless fields and certainly no 4-rank
gauge fields present in this representation. So it seems rather
difficult to envisage solutions with the required central charge. 
Furthermore, solitonic solutions are expected to transform into other
solutions under G, which represents after all a
symmetry of the equations of motion. Thus one would expect the central
charge lattice to eventually cover all possible values consistent with
the group G, and not just with ${\rm H}_{\rm R}$. However, the
supersymmetry algebra itself does not allow for all central charges
consistent with the group G. 

Let us recall that the hidden symetries of the toroidal
compactifications of $d=11$ supergravity \cite{CJS} are realized as 
continuous symmetries upon truncation to the massless modes. When
including the BPS modes, the symmetry can no longer be realized in
this form. The central charge values will constitute a certain
lattice and the hidden symmetry group will be restricted to an
arithmetic subgroup of ${\rm E}_{n(n)}$ that leaves the charge lattice
invariant. It has been conjectured \cite{HullTownsend} that this
arithmetic subgroup, called U-duality, is in fact a symmetry group of
the full M-theory (compactified on a hyper-torus). The fact that, in
low dimensions, the central charges cannot be assigned to representations 
of the hidden symmetry casts some doubt on the assertion that 
${\rm E}_{8(8)}(\bf{}Z)$ and ${\rm E}_{9(9)}(\bf{}Z)$ are symmetries 
of M-theory reduced to three and two dimensions, respectively. 
Similar comments apply to the relevance of ${\rm E}_{7(7)}({\bf Z})$ 
and and ${\rm E}_{6(6)} ({\bf Z})$ for the stringlike and membrane-like
solitonic excitations in $d=4$ and $d=5$, respectively.
In fact, below three dimensions the situation is even less clear,
because the canonical realization of the hidden global symmetry
does not yield the expected affine algebra ${\rm E}_{9(9)}$, but rather
a quadratic (Yangian) algebra over  ${\rm E}_{8(8)}$  \cite{KNS1}.

Another question that poses itself is that, if M-theory does indeed
possess the U-dualities as a symmetry of the full theory, then some
trace of the ${\rm E}_{n(n)}$ symmetry of the compactified theory
could still be present at the level of $d=11$ supergravity. This
expectation is not unreasonable in view of the fact that, while the
towers of Kaluza-Klein states are not in representations of ${\rm
E}_{n(n)}$,  one could envisage completing the theory by adding some
of the missing BPS states in order to regain some (approximate) 
invariance. Hence it is of interest to see whether $d=11$
supergravity, without truncation to the massless states in some
toroidal compactification, has any features reminiscent of
the ${\rm E}_{n(n)}$ symmetry that one finds in the truncated
theories. As we will argue in the next section, this indeed
turns out to be the case.
In section~4 we will then discuss the
possibilities for modifying $d=11$ supergravity in such a way that we
regain (some part of) the  ${\rm E}_{n(n)}$ invariances.

\section{Hidden Symmetries in Eleven Dimensions}

Our discussion of central charges and the way in which they
arise from the $d=11$ ancestor theory raises the question
as to what the role of the hidden ${\rm E}_{n(n)}$ symmetries
is in the context of that theory. As we mentioned already in the
previous section it has been conjectured that an arithmetic subgroup
of the nonlinearly realized ${\rm E}_{n(n)}$ symmetries is 
an exact symmetry of (toroidally compactified)
M-theory. On the other hand, $d=11$ supergravity
as such only exhibits this invariance when truncated to the massless
fields in a toroidal compactification. From the perspective of that
compactification the massive Kaluza-Klein states are responsible for
the fact that the invariance is lost for $d=11$ supergravity. This is
not surprising in view of the fact that, as noted in the previous
section, the Kaluza-Klein states are 
incomplete and do not constitute ${\rm E}_{n(n)}$ multiplets of BPS states.
This observation suggests that it may be possible to extend the theory
in such a way that an arithmetic subgroup of ${\rm E}_{n(n)}$ could become
an exact invariance. We will turn to this question in the next
section, but here we wish to point out that, indeed, full
uncompactified $d=11$ supergravity shows traces of the hidden
symmetries, which suggests that they appear not merely in special
compactifications and corresponding truncations. 
We review the evidence for this idea which was
presented in \cite{dewnic1,nic1} already some time ago, and
discuss it in the light of the more recent developments.
There are intriguing indications of a hidden ``exceptional
geometry'' of $d=11$ supergravity that remains to be discovered
(see \cite{KNS2} for a more recent discussion of this point).
While the work of \cite{dewnic1} was aimed chiefly at establishing
the consistency of the Kaluza-Klein truncation of $d=11$ supergravity
compactified on $S^7$ to its massless sector \cite{dewnic2}, we are thus 
motivated here by the desire to understand the dynamics of the 
non-perturbative M-theory degrees of freedom, and the role
of hidden symmetries in the full M-theory. 

Let us first summarize the main results of \cite{dewnic1,nic1}, where
alternative versions of $d=11$ supergravity were constructed 
with local $\su$ and $\so$ tangent-space symmetries, respectively, 
which are gauge equivalent to the original version of \cite{CJS}. 
This equivalence holds at the level of the equations of
motion\footnote{
  Recall that even the Lagrangian of $N=8$ supergravity
  \cite{CJ} cannot be obtained directly from eleven dimensions; to
  exhibit the hidden symmetries, one must pass through the
  equations of motion.}. In both of these new 
versions the supersymmetry variations acquire a polynomial form from 
which the corresponding formulas for the maximal supergravities 
in four and three dimensions can be read off directly and without 
the need for any duality redefinitions.
This reformulation can thus be regarded as a step towards the 
complete fusion of the bosonic degrees of freedom of $d=11$ 
supergravity (i.e. the elfbein $E_M^{~A}$ and the antisymmetric 
tensor $A_{MNP}$) in a way which is in harmony with the hidden 
symmetries of the dimensionally reduced theories.

For lack of space we restrict attention to the bosonic 
sector, and first describe the version of \cite {dewnic1}
exposing a hidden ${\rm E}_{7(7)}$ structure of $d=11$ supergravity.
There is also a version involving $E_{8(8)}$ \cite{nic1}, but
in order not to overburden the notation with too many different 
kinds of indices, we will summarize the pertinent results separately 
in the second part of this section. At any rate, readers are 
advised to consult the original papers for further explanations
and the more technical details of the construction.

The first step in the procedure is to break the original tangent-space
symmetry SO(1,10) to its subgroups $\SB$ and $\sa$, respectively, 
through a partial choice of gauge for the elfbein. In a second step,
one enlarges these symmetries again to $\su$ and $\so$ by introducing 
new gauge degrees of freedom. This symmetry enhancement requires 
suitable redefinitions of the bosonic and fermionic fields, and
their combination into tensors with respect to the new tangent-space
symmetries.  
The construction thus requires 4+7 and 3+8 splits of the $d=11$ 
coordinates and indices, respectively, implying a similar split 
for all tensors of the theory. It is important, however, that the 
dependence on all eleven coordinates is retained throughout. The
alternative theory remains fully equivalent to the original formulation
of \cite{CJS} upon suitable gauge choices. 

The elfbein and the three-index photon are thus combined into 
new objects covariant with respect to the new tangent-space symmetry. 
In the special Lorentz gauge preserving either $\SB$ or $\sa$ the 
elfbein takes the form
\bea
E_M^{~A} = \left(\begin{array}{cc} 
            \Delta^{-s}e_\mu^{~a} & B_\mu^{~m} e_m^{~a}\\[2mm]
            0& e_m^{~a}   \end{array} \right)\;,
\label{11bein}
\eea
where curved $d=11$ indices are decomposed as $M=(\mu ,m)$ with 
$\mu =0,1,2(,3)$ and $m= (3,)4,...,10$, with a similar decomposition 
of the flat indices; furthermore, $\Delta := {\rm det} \, e_m^{~a}$ and
$s= 1/(d-2)$ for $d=4$ and 3, respectively.
In this gauge, the elfbein contains the (Weyl rescaled) 
drei- or vierbein and the Kaluza-Klein vectors $B_\mu{}^{m}$. 
The internal vielbein is replaced by a new object, which we refer to as a
generalized vielbein, and which can be identified by a careful analysis
of the supersymmetry variation of $B_\mu{}^m$. For the 4+7 split, this
generalized vielbein, denoted by $e^m_{AB}$, carries an {\it upper} internal 
world index, and has lower SU(8) indices $A,B=1,\ldots,8$. It is
anti-symmetric in the indices $A,B$ so that it transforms in the $\bf
28$ representation. Explicitly,
\beq
e^m_{AB}\rightarrow U_A{}^C \,U_B{}^D \,e^m_{CD}\,,
\eeq
where $U_A{}^C$ is an SU(8) matrix depending on all eleven 
coordinates. 
By including its complex conjugate,
\beq
e^{mAB} := (e^m_{AB})^*\,,
\eeq
the generalized vielbein is thus given by
the complex tensor $(e^{mAB} , \,e^m_{AB})$, which, for given $m$,
constitutes the ${\bf 56}$ 
(pseudo-real) representation of ${\rm E}_{7(7)}$; according to its
maximal compact subgroup SU(8) this representation branches into ${\bf
28}\oplus  \overline{\bf 28}$ 

The generalized vielbein contains the original
siebenbein $e_m^{~a}$, as can be seen by choosing a special 
SU(8) gauge such that
\beq
e^m_{AB} := i \Delta^{-1/2} e_a{}^m \G^a_{AB}\,, \label{bare-bein}
\eeq
where $m=4,...,10$ and $\G^a$ are the standard SO(7) $\G$-matrices
(our conventions are such that $e^m_{AB}$ is real in this gauge).
Being the inverse densitized internal siebenbein contracted with 
an SO(7) $\Gamma$-matrix, our generalized vielbein object is very 
much analogous to the inverse densitized triad in Ashtekar's 
reformulation of Einstein's theory \cite{A}.

The generalized vielbein has many more components than the original 
siebenbein, but of course the number of physical degrees of freedom
is the same as before. Some of the redundant degrees of freedom
are taken care of by the SU(8) gauge symmetry, but further algebraic
constraints must exist to match the original physical content
of the theory. These constraints are indeed present and can be derived
by making use of properties of SO(7) $\Gamma$-matrices. 
An obvious one is the ``Clifford property'', already identified in
\cite {dewnic1}:
\beq
e^m_{AC}\, e^{nCB} + e^n_{AC}\, e^{mCB} = \ft14 \delta_A^B
\,e^m_{CD}\, e^{nDC}\,.  
\eeq
{}Furthermore, from \eqn{bare-bein} one obtains a formula for the
original seven-metric, 
\beq
({\rm det} \, g)^{-s} g^{mn} = \ft18 \,  e^m_{CD} \,e^{nCD} \,,
\eeq
in terms of the new generalized vielbein, which immediately
yields the ``master formula'' for the full non-linear metric ansatz 
in the Kaluzu-Klein reduction of $d=11$ supergravity. This 
formula has been exploited in recent work on the AdS/CFT 
correspondence, see e.g. \cite{ADSCFT}.

The Clifford property is by itself not enough to reduce
the number of physical degrees of freedom to the desired one. Rather,
it is part of the following set of ${\rm E}_{7(7)}$ covariant constraints,
\bea
e^m_{AB} e^{nAB} -e^n_{AB} e^{mAB} &=& 0 \,,\nonumber \\
e^m_{AC} e^{nCB} + e^n_{AC} e^{mCB} - 
     \ft14 \delta_A^B e^{m}_{CD} e^{nDC} &=& 0 \,, \nn
e^m_{[AB} e^n_{CD]} - \ft1{24} \varepsilon_{ABCDEFGH}
   e^{mEF} e^{nGH} &=& 0\,. 
\eea
These equations correspond to the singlet and the $\bf 133$ 
in the ${\rm E}_{7(7)}$ decomposition, 
\beq 
{\bf 56} \otimes  {\bf 56}
\rightarrow {\bf 1} \oplus{\bf 133} \oplus{\bf 1463} \oplus{\bf 1539}\;.
\eeq
The constraints can thus be rephrased as the statement that
the product $e^m \otimes  e^n$ only contains the $\bf 1463$ and
$\bf 1539$ representations of ${\rm E}_{7(7)}$. 

In addition to the algebraic constraints, the generalized vielbein
satisfies a set of differential relations, called the ``generalized 
vielbein postulate'' in \cite{dewnic1}. In order to state them, 
we need suitable ${\rm E}_{7(7)}$ connections ${Q_M}^{\!A}{}_{\!B}$ and 
$P_M^{ABCD} = \ft1{24} \varepsilon^{ABCDEFGH} P_{M\, EFGH}$
in {\em eleven} dimensions. These are built out of the SO(1,10)
coefficients of anholonomity and the four-index field strength 
$F_{MNPQ}$ of $d=11$ supergravity in the way explained in \cite{dewnic1}; 
since the explicit expressions are somewhat cumbersome we refer readers
there for details. The vector ${Q_M}^{\!A}{}_{\!B}$ acts as the
connection for the local SU(8) transformations and is therefore in the
$\bf 63$ representation of that group. The tensor $P_M^{ABCD}$ transforms as
the selfdual $\bf 35$ under the action of SU(8). Together they
constitute the (adjoint) $\bf 133$ representation of ${\rm E}_{7(7)}$. 
For the massless theory these quantities are directly
related to the pull-backs to $d=11$ space-time of the tangent-space
connection and vielbein associated with the homogeneous space 
${\rm E}_{7(7)}/{\rm SU}(8)$. 
 
The generalized vielbein postulate takes the form 
\bea
\DC_\mu e^m_{AB} + \ft12 \DC_n {B_\mu}^n  e^m_{AB} + \DC_n {B_\mu}^m  e^n_{AB}
+ 2\, Q_{\mu[A}{}^{\!C} e^m_{B]C} + P_{\mu ABCD}\, e^{mCD} &=& 0\,,  \nn
\DC_n e^m_{AB}
+ 2\, Q_{n[A}{}^{\!C} e^m_{B]C} + P_{n ABCD}\, e^{mCD} &=& 0\,, 
\eea
where
\beq\label{Dmu}
\DC_\mu := \partial_\mu - {B_\mu}^m \DC_m\,,
\eeq
for $\mu=0,1,2,3$ and 
\bea 
\DC_m e^n_{AB} &:=& \partial_m e^n_{AB} + {\Gamma_{mp}}^n e^p_{AB}
                  + \ft12 {\Gamma_{mp}}^p  e^n_{AB}\,,\nn
\DC_m {B_\mu}^n &:=& \partial_m  {B_\mu}^n + {\Gamma_{mp}}^n
{B_\mu}^p\,, 
\eea 
for the internal indices. The extra term with ${\Gamma_{mp}}^p$ in
the above relation arises because the generalized vielbein 
transforms as a density. Observe
that the affine connection ${{\G}_{mn}}^p$ still depends on
all eleven coordinates, and that the covariance of these relations under 
general internal coordinate transformations (with parameters $\xi^m(x,y)$) 
had not been previously exhibited in \cite{dewnic1} where the generalized 
vielbein postulate was given without the affine connections. 
The relevant extra terms are obtained by uniformly replacing
\beq  
e_a{}^{\!n} \partial_m e_{nb} \rightarrow
 e_a{}^{\!n} \left(\partial_m e_{nb} - {\G_{mn}}^p e_{pb}\right)\,,
\eeq
in the relevant formulas of \cite{dewnic1,nic1} defining the connection. 
For the explicit verification of the above relations with the fully 
covariant derivatives, it is furthermore useful to observe that
the relevant terms appearing in the connection coefficients 
$Q_\mu$ and $P_\mu$ contain the combination
$$
e_a^{~m} \left( \partial_m {B_\mu}^n e_{nb} + 
               {B_\mu}^n \partial_n e_{mb} \right) =
e_a^{~m} \big( \partial_m ({B_\mu}^n e_{nb}) + 
              2 {B_\mu}^n \partial_{[n} e_{m]b} \big)\,,
$$
which remains unchanged if we replace $\partial_m$ by $\DC_m$,
provided the affine connection is torsion-free.

We emphasize that the 
affine connection is still arbitrary at this point, as it cancels 
between the different terms in the generalized vielbein postulate.
A convenient choice is the standard Christoffel connection,
which is obtained by setting
\beq
\DC_m (\Delta^{-1} g^{np}) = 2 P_m^{ABCD}\, e^n_{AB} \,e^p_{CD} =
0\,. 
\eeq

So we conclude that all the quantities introduced above comprise ${\rm
E}_{7(7)}$ representations. We stress once more that we are still dealing
with the full $d=11$ supergravity theory. This pattern continues. For
instance, the supersymmetry variation of the generalized vielbein
takes a form that closely resembles the four-dimensional
transformation rule for the massless modes (in the truncation to the
massless modes, $e^m_{AB}$ is proportional to the ${\rm
E}_{7(7)}/{\rm SU}(8)$ coset representative),
\beq
\delta e^m_{AB} =  - \sqrt{2}\,\Sigma_{ABCD} \, e^{m\,CD}\,,
\eeq
where
\beq
\Sigma_{ABCD}=\bar \e_{[A} \chi_{BCD]} + \ft1{24} \,\e^E\chi^{FGH}\,,
\eeq
where $\e_A$ and $\chi_{ABC}$ denote the supersymmetry parameters and
the spin-1/2 fields, respectively. It takes a little more work 
to check that the covariantizations with respect to ${\G_{mn}}^p$
introduced 
above do not alter the form of the supersymmetry variations given in
\cite{dewnic1}, except for extra terms involving  ${\G_{mn}}^n$ 
necessary because the redefined supersymmetry transformation parameter 
is also a density of weight $\frac14$ (the original supersymmetry
parameter is of zero weight). Similarly, the bosonic and fermionic 
equations of motion can be cast into a fully SU(8) covariant form.

In spite of the fact that the theory can be formulated elegantly in
terms of ${\rm E}_{7(7)}$ quantities, it cannot be invariant under 
${\rm E}_{7(7)}$. The obvious reason for that is the presence of the
Kaluza-Klein gauge fields $B_\m{}^m$, which do not constitute a proper
representation. However, when restricting ourselves to the massless
modes in the toroidal representations, these fields disappear in the
generalized vielbein postulate. On the other hand, the inability to
preserve ${\rm E}_{7(7)}$ invariance for these fields is precisely
related to the deficiencies in the central charge assignments that we
discussed in the previous section. If this deficiency could somehow be
lifted, then it might be possible to regain the hidden symmetry with
BPS states present.  

For the 3+8 split one has analogous results \cite{nic1,NM,KNS2}, but
with a different decomposition of indices. The hidden  symmetry 
of the theory is ${\rm E}_{8(8)}$ \cite{Julia,MS}
with a local SO(16) replacing the SU(8) of the 4+7 split. The SO(16)
vector representation $\bf 16$ is labeled by indices $I,J
=1,\ldots,16$, while $A,B= 1,\ldots ,128$ 
labels the ${\bf 128}_s$ chiral spinor and 
$\dot A,\dot B= 1,\ldots ,128$  the ${\bf 128}_c$ opposite-chirality
spinor of SO(16). The SO(8) tangent
space group of the original $d=11$ theory is embedded into SO(16)
according to ${\bf 16}_v \rightarrow {\bf 8}_s
\oplus {\bf 8}_c$. The two spinor representations then branch according
to ${\bf 128}_s \rightarrow ({\bf 8}_v \otimes {\bf 8}_v) \oplus
({\bf 8}_s \otimes {\bf 8}_c)$, 
and according to ${\bf 128}_c \rightarrow ({\bf 8}_v \otimes {\bf
8}_c) \oplus ({\bf 8}_s \otimes {\bf 8}_v)$. The SO(16)
adjoint representation $\bf 120$ and the spinor representation ${\bf
128}_s$ (or its adjoint) constitute the $\bf 248$ 
representation of ${\rm E}_{8(8)}$, where we remind the reader of the
well-known fact (relevant below) that the adjoint and the fundamental
representation of this group coincide. Hence the $\bf 248$
representation can be labeled by the indices $\AC=([IJ],A)$ according
to the SO(16) decomposition ${\bf 120} \oplus {\bf 128}$.  
Observe that the SO(16) index pairs $[IJ]$ correspond to the SO(8)
index pairs $[\a\b]$, $\a\dot\b$ and $[\dot\a\dot\b]$; the SO(16)
spinor indices $A$ and $\dot A$ also correspond to SO(8) index pairs,
namely to $ab$ and $\a\dot\b$, and to $a\a$ and $a\dot\a$, respectively. 
Here the SO(8) indices $a,\A , \dot\A$ label the ${\bf 8}_v, {\bf
8}_s$ and ${\bf 8}_c$ representations of SO(8), respectively.

The matter-like bosonic degrees of freedom are now combined into a
generalized 
vielbein $e^m_\AC \equiv (e^m_{IJ},e^m_A )$ on which local SO(16)
acts reducibly according to
\beq
e^m_\AC \rightarrow {U_\AC}^\BC e^m_\BC\,,
\eeq
with ${U_\AC}^\BC$ in the ${\bf 120} \oplus {\bf 128}$ representation. 
As before, we can relate this new vielbein to the original achtbein
in a special gauge. 
In order to avoid introducing yet 
more notation we refer readers to \cite{nic1} for details, and 
simply quote the result, 
\bea
(e^m_{IJ},e^m_A ) := \left\{ \begin{array}{ll}
     \Delta^{-1} e_a^{~m} \Gamma^a_{\A \dot \B} 
            & \mbox{if $[IJ]$ or $A = \A \dot \B$ \,,}\\
       0 & \mbox{otherwise\,.}
       \end{array} \right.  \label{bare-bein-8}
\eea

As before, the ${\rm E}_{8(8)}$ vielbein $e^m_\AC$ has far more components
than there are physical degrees of freedom, and therefore must again
satisfy a number of algebraic constraints \cite{NM}. We have
\bea
e^m_A \,e^n_A - \bruch{1}{2}e^m_{IJ}\, e^n_{IJ} = 0 \,,\label{id1}
\eea
and
\bea
\G^{IJ}_{AB} \Big( e^m_B\, e^n_{IJ} - e^n_B \,e^m_{IJ} \Big) = 0\,,
\qquad 
\G^{IJ}_{AB}\, e^m_A \,e^n_B + 4 e^m_{K[I}\, e^n_{J]K} = 0 \,,\label{id2}
\eea
where $\Gamma^I_{A\dot A}$ are the standard SO(16) $\G$-matrices and 
$\G_{AB}^{IJ}\equiv (\G^{[I} \G^{J]})_{AB}$; the minus 
sign in (\ref{id1}) reflects the fact that we are dealing with
the maximally non-compact ${\rm E}_{8(8)}$. Obviously, (\ref{id1}) 
and (\ref{id2}) correspond to the singlet and the adjoint 
representations of ${\rm E}_{8(8)}$ in the product $e^m\otimes e^n$.
More complicated are the following relations transforming in the 
$\bf 3875$ representation of ${\rm E}_{8(8)}$, 
\bea
e^{(m}_{IK} \,e^{n)}_{JK} - \bruch{1}{16} \delta_{IJ} \,
e^m_{KL}\, e^n_{KL}  &=& 0 \,,  \nn
\G^K_{\dot A B}\, e^{(m}_B \,e^{n)}_{IK} - \bruch{1}{14}
\G^{IKL}_{\dot A B}\, e^{(m}_B \, e^{n)}_{KL} &=& 0 \,,\nn
e^{(m}_{[IJ}\, e^{n)}_{KL]} + \bruch{1}{24}
e^m_A \,\G^{IJKL}_{AB} \,e^n_B  &=& 0\,. \label{id4}
\eea
These relations can be elegantly summarized by means of ${\rm E}_{8(8)}$
projectors \cite{KNS2}
\beq
{(\PC_{\rm j})_{\AC\BC}}^{\CC\DC} e^m_{~\CC} e^n_{~\DC} = 0\,,
\eeq
for the ${\bf j} ={\bf 1}, {\bf 248}$ and ${\bf 3875}$ representations
appearing in the product,
\beq
{\bf 248} \otimes {\bf 248} \rightarrow
{\bf 1} \oplus {\bf 248}\oplus {\bf 3875} \oplus {\bf 27000}
\oplus {\bf 30380}\;.
\eeq
In comparison with the 4+7 case we note the appearance
of a fifth representation in this decomposition corresponding to
$\bf 3875$ which has no analog in $E_{7(7)}$.

For the differential relations we again need composite connections
that belong to the Lie algebra of ${\rm E}_{8(8)}$. These have components
\beq
\QC_M{}^\AC \equiv (Q_{M}{}^{IJ}, P_{M}{}^{A})\,,
\eeq 
whose explicit expressions in terms of the $d=11$ coefficients of 
anholonomity and the four-index field strength $F_{MNPQ}$ can 
be found in \cite{nic1}. The generalized vielbein postulate now takes 
an even simpler form: with the ${\rm E}_{8(8)}$ structure constants
${f_{\AC\BC}}^\CC$, we have
\bea
\DC_\mu e^m{}_{\!\AC} + \DC_n {B_\mu}^n \, e^m{}_{\!\AC} + \DC_n
{B_\mu}^m \, e^n{}_{\!\AC}
+ {f_{\AC\BC}}^\CC \QC_\mu{}^\BC e^m{}_{\!\CC} &=& 0\,,\nn
\DC_n e^m{}_\AC
+ {f_{\AC\BC}}^{\!\CC} \QC_n{}^\BC e^m{}_{\!\CC} &=& 0\,,
\eea
where $\DC_\mu$ and $\DC_m$ are same as before, except for the fact
that the generalized vielbein transforms as a density of different
weight as compared to the case $d=4$ (cf. (\ref{bare-bein}) and
(\ref{bare-bein-8})).  
Like (\ref{id1})--(\ref{id4}), the differential relations are thus 
covariant under general coordinate transformations as well as ${\rm
E}_{8(8)}$. Just as before, we note that the full theory does not 
respect ${\rm E}_{8(8)}$ invariance. 

The new versions yield the maximal supergravities in $d=4$ and $d=3$
directly and without further ado. In particular, one obtains in this way
the SU$(8)$ and SO$(16)$ covariant equations of motion, which combine
equations of motion and Bianchi identities of the original theory in
eleven dimensions \cite{dewnic1}. Furthermore, the reduction of $d=11$ 
supergravity to three dimensions yields $d=3, N=16$ supergravity \cite{MS}, 
and is accomplished rather easily, since no duality redefinitions 
are needed any more, unlike in \cite{CJ}. The propagating bosonic 
degrees of freedom in three dimensions are all scalar, and combine 
into a matrix $\VC(x)$, which is an element of a non-compact 
${\rm E}_{8(8)}/\s$ coset space, and whose dynamics is governed by a 
non-linear $\sigma$-model coupled to $d=3$ gravity. The identification 
of the 248-bein with the $\sigma$-model field $\VC\in {\rm E}_{8(8)}$ is
given by  
\bea 
e^m_{IJ}(x) = \bruch{1}{60}{\rm Tr} \, \big[ Z^m\, \VC(x)\, X^{IJ}\,
\VC^{-1}(x)  \big] \,,\qquad
e^m_A(x) = \bruch{1}{60}{\rm Tr} \, \big[ Z^m \,\VC(x)\, Y^A\, \VC^{-1}(x) 
\big]\,, 
\label{bein1}
\eea
where $X^{IJ}$ and $Y^A$ are the compact and non-compact 
generators of ${\rm E}_{8(8)}$, respectively, and where the $Z^m$ for 
$m=3,...,10$ are eight commuting nilpotent generators (hence obeying 
${\rm Tr} (Z^m Z^n) = 0$ for all $m$ and $n$). The verification of
these assertions, and in particular of (\ref{id4}), relies on the
very special properties of the ${\rm E}_{8(8)}$ Lie algebra (we refer
to \cite{KNS2} for details). 

A very interesting aspect of the 3+8 split is that the above relations
can be exploited to argue that the ${\rm E}_{8(8)}$ matrix $\VC$ is
present already in eleven dimensions and thus depends on the
eleven-dimensional coordinates \cite{KNS2}. Namely, the fundamental and 
adjoint repesentations of ${\rm E}_{8(8)}$ being the same, we have
\beq
\VC^{-1}(x,y)\, t^\AC \,\VC(x,y) = {\VC^\AC}_{\!\BC}(x,y) \,t^\BC
\eeq
for all ${\rm E}_{8(8)}$ generators $t^\AC$. Hence,
\beq
e^m{}_{\!\AC}(x,y) = {\VC^m}_{\!\!\AC}(x,y)\,. 
\eeq
Put differently, the generalized vielbein (as a rectangular
matrix) is a submatrix of a full 248-bein in eleven dimensions!
This suggests that we should enlarge the range of indices $m$
to run over the whole group $E_{8(8)}$, with a corresponding 
increase in the number of dimensions, which, as we will see, is also
suggested by the analysis of central charges. A first step in this
direction was taken in \cite{KNS2}, where it was shown that the
28 further vector fields originating from the three-form potential 
of $d=11$ supergravity necessitate the introduction of 28 further
components $e_{mn\AC}$, thereby enlarging the range of indices to 36.

Evidently many of the formulas displayed above are trivially satisfied
for toroidal compactifications --- for instance, all the connection
components in the internal dimensions simply vanish in the truncation
to the massless sector. One would therefore like to check the above 
results in the context of compactifications of $d=11$ supergravity
on non-trivial internal manifolds, as such compactifications can
provide valuable ``models'' for the exceptional geometry that we have
been alluding to. To date there is only one model of this type, namely
the $AdS_4\times S^7$ compactification of $d=11$ supergravity \cite{dewnic2} 
(the $AdS_7\times S^4$ truncation of \cite{PvN} could eventually provide 
another model, but those results remain to be analyzed from the 
point of view taken here). In that case, the internal connection
components ${Q_m}^{\!A}{}_{\!B}$ and $P_m^{ABCD}$ do survive the truncation
to the massless modes and are metamorphosed into the ``$T$ tensor''
describing the couplings of the scalars and the fermions in gauged
supergravity \cite{dewnic3,dewnic2}. This is related to the remarkable fact 
that for gauged $N=8$ supergravity in four and five dimensions,
this $T$-tensor actually transforms as a representation of the 
${\rm E}_{n(n)}$ symmetry group (for $n=7$ and $n=6$, respectively), even
though ${\rm E}_{n(n)}$ is no longer a symmetry of the gauged theory!
More specifically, for $d=4$ we have \cite{dewnic4}
\beq
{\bf 36} \oplus {\bf 420} + c.c. \quad \Big( {\rm of} \; {\rm SU}(8) \Big)
\longrightarrow {\bf 912}  \quad \Big( {\rm of} \; {\rm E}_{7(7)} \Big)\,,
\eeq
while for $d=5$ \cite{GRW}
\beq
{\bf 36} \oplus {\bf 315} \quad \Big( {\rm of} \; {\rm USp}(8) \Big)
\longrightarrow {\bf 351}  \quad \Big( {\rm of} \; {\rm E}_{6(6)}
\Big)\,. 
\eeq
The analogous decomposition for $d=3$, namely 
\beq\label{E8T}
{\bf 135} \oplus {\bf 1820} \oplus {\bf 1920} \quad \Big( {\rm of} \; {\rm
SO}(16) \Big) 
\longrightarrow {\bf 3875}  \quad \Big( {\rm of} \; {\rm E}_{8(8)} \Big)\,,
\eeq
has recently been invoked to construct a gauged maximal supergravity 
in three dimensions \cite{NS}. There, the consistency of the gauged
theory is imposed by searching for gauge groups such that the
$T$-tensor admits precisely the decomposition (\ref{E8T}), thereby
turning the derivation of \cite{dewnic3} upside down.

\section{BPS-extended Supergravity}

In the foregoing section we reviewed the evidence for hidden ${\rm
E}_{7(7)}$  and ${\rm E}_{8(8)}$ structures of $d=11$
supergravity. However, the fact 
that certain sectors of the theory assume an ${\rm E}_{n(n)}$ covariant 
form does not mean that ${\rm E}_{n(n)}$ is actually a symmmetry of  $d=11$ supergravity, as is already obvious from the fact
that the vector fields ${B_\mu}^m$ in the formula (\ref{Dmu}) 
cannot be assigned to representations of ${\rm E}_{n(n)}$.
Rather, our results should be interpreted as an indication that
there exist extensions of $d=11$ supergravity which do possess
these symmetries. Recent developments in string theory have led
to the conjecture that the so-called U-duality 
group, which is an integer-valued subgroup of the nonlinearly 
realized ${\rm E}_{n(n)}$ symmetries is actually an exact symmetry of
(toroidally compactified) M-theory and therefore acts on the BPS 
states as well (see e.g. \cite{ObersPioline} for a recent review). 
It is therefore not unreasonable to conjecture the existence of yet 
larger theories unifying the BPS degrees of freedom. We will refer 
to the effective field theories incorporating all these degrees 
of freedom as ``BPS extended supergravities''.

The existence of central charges other than those associated with 
the momentum states on the hyper-torus strongly suggests that there
are extra space-time dimensions that would be similarly associated 
with the remaining central charges. Such an extension would in 
particular imply the existence of further Kaluza-Klein vector fields in
(\ref{Dmu}), which would then couple to the non-momentum central
charges. In most compactifications one has precisely the right number
of vector gauge fields. In five dimensions there is one singlet
pointlike central charge without a corresponding gauge field; in four
dimensions there are 28 gauge fields and 56 central charges. The
extra 28 charges are associated with monopoles and here the charges
are mutually nonlocal in the sense that one cannot incorporate
electric and magnetic charges simultaneously in a local field
theory. Admittedly this may be an obstacle in associating extra 
dimensions to all the 56 central charges. The idea of introducing   
extra dimensions for the description of supersymmetric theories 
with central charges is, in fact, an old one \cite{Sohnius}, but we 
believe that the results reviewed above make it even more compelling.
We would therefore expect there to be bigger theories living in 
4+56 dimensions (for the  ${\rm E}_{7(7)}$ version of \cite{dewnic1}), 
and 3+248 dimensions (for the ${\rm E}_{8(8)}$ version of \cite{nic1}).
However, it remains to be seen whether these extra dimensions are
really on the same footing as ordinary space-time dimensions.
In this section, we retreat a little in order to explore the
implications of this idea in the context of a simpler example, 
namely $d=9$ supergravity, where there are only three central
charges (cf. table~\ref{CC1}). In that case, one has 
not only a detailed understanding of certain BPS states, contained in
so-called KKA and KKB supermultiplets \cite{ADLN}, but there is 
also a candidate theory whose effective field theory description
contains both the supergravity and the BPS degrees of freedom, as 
well as the massive IIA and IIB superstring degrees of freedom,
and would therefore serve as a prime example of the BPS-extended
supergravity that we have in mind here, namely the supermembrane 
\cite{BST}! This point was, in fact, already made in \cite{JHS} 
where it was first proposed to view the Kaluza-Klein and the winding
states on the M-theory torus as manifestations of an underlying
supermembrane theory. In dimensions less than nine, yet more
(pointlike) BPS degrees of freedom  
will arise. This suggests the existence of yet bigger theories 
``beyond the supermembrane'' which eventually would also account for
the states associated with the five-brane charges.

The compactification of M-theory on a torus $T^2$ is expected to 
comprise both IIA and IIB superstring theory, and is hence conjectured 
to be invariant under the S-duality group SL$(2,{\bf Z})$. In ten
spacetime dimensions the massive supermultiplets of IIA  
and IIB string theory coincide, whereas the massless states comprise 
inequivalent supermultiplets, for the simple reason that they transform 
according to different representations of the SO(8) helicity group.
When compactifying the theory on a circle, massless states IIA and
IIB states in nine spacetime dimensions transform according to 
identical SO(7) representations of the helicity group and constitute 
equivalent supermultiplets. The corresponding interacting field theory 
is the unique $N=2$ supergravity theory in nine spacetime dimensions. 
However, the supermultiplets of the BPS states, which carry momentum
along the circle, remain inequivalent, as they remain assigned to
the inequivalent representations of the group SO(8) which is now
associated with the restframe (spin) rotations of the massive states. 
The Kaluza-Klein states of the IIA theory constitute the so-called KKA
supermultiplets, whereas those of the IIB theory constitute the
(inequivalent) KKB multiplets. It was proven in \cite{ADLN} that 
the IIA winding states constitute KKB supermultiplets and the 
IIB winding states constitute KKA supermultiplets, so that T-duality
remains valid. 

One can obtain the same result for the
eleven-dimensional (super)membrane, which contains excitations
corresponding to all the BPS states found in the supermultiplet 
analysis. Assuming that the two-brane charge
takes values only in the compact coordinates labeled by 9 and 10, 
which can be generated by wrapping the membrane over the 
corresponding $T^2$, one readily finds the following expression
for the most general scalar central charge ($\sigma_{1,3}$ are the
real Pauli matrices and the indices $i,j=1,2$ label the two
supersymmetries),   
\beq 
Z^{ij} = Z_{9\,10} \,\d^{ij} - (P_9 \,\sigma_3^{ij} - P_{10} \,
\sigma_1^{ij})\,.
\eeq
This result is, of course, in agreement with the $d=9$ entry of table
1 in section 1.  
Here $P_9$ and $P_{10}$ denote the Kaluza-Klein momenta, while $Z_{9\,10}$
is the winding number of the membrane on $T^2$.
Assuming that we are compactifying over a torus with modular parameter
$\tau$ and area $A$, the mass formula takes the form  
\bea
M_{\rm BPS} &=& \sqrt{P_9^{\,2} + P_{10}^{\,2} } + \vert
Z_{9\,10}\vert \nonumber\\
&=& {1\over \sqrt{A\,\tau_2}} \vert q_1 + \tau \,q_2\vert + T_{\rm m}
A \,\vert p\vert \,. \label{BPS-membrane}
\eea
Here $q_{1,2}$ denote the momentum numbers on the torus and $p$ is the
number of times the membrane is wrapped over torus; $T_{\rm m}$
denotes the supermembrane tension. 
Clearly the KKA states correspond to the momentum modes while
the KKB states are associated with the wrapped membranes on
$T^2$. Hence there is a rather natural way to describe the IIA and IIB 
momentum and winding states starting from a (super)membrane in eleven
space-time dimensions \cite{JHS,ADLN}. 

To really construct the BPS-extended supergravity theory associated
with the supermembrane compactified on $T^2$, 
one would have to consider $N=2$ supergravity in nine space-time
dimensions and couple it to the simplest BPS supermultiplets
corresponding to the KKA and KKB states. Nine-dimensional supergravity 
has precisely three gauge fields that couple to the three central charges
discussed above. From the perspective of eleven-dimensional supergravity
compactified on $T^2$ the KKA multiplets are the Kaluza-Klein states. Their
charges transform obviously with respect to an SO(2) associated with
rotations of the coordinates labeled by 9 and 10. Hence we have a
double tower of these charges with corresponding 
KKA supermultiplets. On the other hand from a IIB perspective,
compactified on $S^1$, the KKB states are the Kaluza-Klein states and their
charge is SO(2) invariant. Here we have a single tower of KKB
supermultiplets. From the perspective of nine-dimensional
BPS-extended supergravity one should be able to couple 
both towers of KKA and KKB supermultiplets simultaneously,
thereby arriving at a theory that contains ten-dimensional IIA and 
IIB theories in certain decompactification limits, as well as 
eleven-dimensional supergravity. The BPS-extended theory is
in some sense truly twelve-dimensional with three compact coordinates,
although there is no twelve-dimensional Lorentz invariance, not even 
in a uniform decompactification limit, as the fields never depend on all 
twelve coordinates! Whether this kind of BPS-extended supergravity
offers a viable scheme in a more general context than the one we
discuss here, remains to be seen. Very little work has been done 
in incorporating BPS and/or Kaluza-Klein multiplets into the field
theory. Nevertheless  
in the case at hand we know a lot about these couplings from our 
knowledge of the $T^2$ compactification of eleven-dimensional 
supergravity and the $S^1$ compactification of IIB supergravity. 

For the convenience of the reader we have listed the fields of 
nine-dimensional $N=2$ supergravity listed in
table~1, where we also indicate their relation with the fields of
eleven-dimensional and ten-dimensional IIA/B supergravity upon
dimensional reduction. It is not necessary to work out all the 
nonlinear field redefinitions here, as the corresponding fields can be
uniquely identified by their scaling weights under SO(1,1), a symmetry
of the massless theory that emerges upon dimensional reduction and is 
associated with scalings of the internal vielbeine. 

\begin{table}
\vspace{2mm}
\begin{center}
\begin{tabular}{ccccc}
\hline $D=11$ & IIA & $D=9$ & {~}\quad IIB \quad{~} & SO(1,1) \\ 
\hline\\[-3mm] 
$\hat G_{\mu \nu}$ & $G_{\m \n}$  & $g_{\m \n}$  & $G_{\m \n}$ & $0$
\\[1mm] 
\hline\\[-3mm] 
$\hat{A}_{\mu {\scriptscriptstyle \,9\,10}}$  & $C_{\mu
{\scriptscriptstyle\,9}}$ &  $B_{\m}$ & $G_{\m {\scriptscriptstyle\,9}}$
& $-4$\\[1mm] 
\hline\\[-3mm] 
$\hat G_{\mu {\scriptscriptstyle\,9}}$, $\hat G_{\mu
{\scriptscriptstyle\,10}}$  & $G_{\mu {\scriptscriptstyle\,9}}$ ,
$C_{\mu}$  & $A_{\mu}^{\, \alpha}$ & $A_{\m
{\scriptscriptstyle\,9}}^{\, \alpha}$ & $3$  \\[1mm]  
\hline\\[-3mm] 
$\hat{A}_{\m \n {\scriptscriptstyle\,9}}$, $\hat{A}_{\m \n
{\scriptscriptstyle\,10}}$ & $C_{\m \n {\scriptscriptstyle\,9}}, C_{\m
\n}$ &  $A_{\m\n}^{\,\a}$ &  $A_{\m\n}^{\,\a}$  & $-1$  \\[1mm]
\hline\\[-3mm] 
$\hat A_{\m \n\rho}$  & $C_{\m \n \rho}$  & $A_{\m\n\rho}$  &
$A_{\m\n\rho\sigma}$ & 2  \\[1mm] 
\hline\\[-3mm] 
$\hat G_{{\scriptscriptstyle 9 \,10}}$, $\hat{G}_{{\scriptscriptstyle9\,
9}}$, $\hat G_{{\scriptscriptstyle 10\,  10}}$ & \quad $\phi$,  
$G_{{\scriptscriptstyle 9\,9}}$, $C_{\scriptscriptstyle  9}$ \quad&
$\left\{ \begin{array}{l} \phi^\a  \\ \exp (\sigma ) \end{array}
\right. $ & $\begin{array}{l} \phi^\a \\ G_{{\scriptscriptstyle
9\,9}} 
\end{array}$ & $\begin{array}{r} 0 \\  7 \end{array}$
\\[4mm] 
\hline 
\end{tabular}
\vspace{.3cm}
\caption{The bosonic fields of the eleven dimensional, type-IIA,  
nine-dimensional $N=2$ and type-IIB  supergravity theories. 
The eleven-dimensional and ten-dimensional indices, respectively, are 
split as $\hat{M} =(\mu, 9,10)$ and $M=(\mu,9)$, where $\mu = 0,1,\ldots 8$. 
The last column lists the SO(1,1) scaling weights of the fields.}
\end{center}

\end{table}

Of particular relevance are the three abelian vector gauge
fields. There are the two vector fields $A_\m^\a$, which are the
Kaluza-Klein 
photons of the $T^2$ reduction of eleven-dimensional supergravity
and which couple therefore to the KKA states. From the IIA perspective
these correspond to the Kaluza-Klein states on $S^1$ and the D0
states. From the IIB side they 
originate from the tensor fields, which confirms that they couple to
the IIB (elementary and D1) winding states. These two fields transform 
under SL(2), which can be understood {from} the perspective of the
modular transformation on $T^2$ as well as of the S-duality
transformations that rotate the elementary with the D1 strings.
The third gauge field, denoted by $B_\m$, is an SL(2)
singlet and is the Kaluza-Klein photon on the IIB side, so that
it couples 
to the KKB states. On the IIA side it originates from the IIA tensor
field, which is consistent with the fact that the IIA winding states
constitute KKB supermultiplets.

The resulting BPS-extended theory incorporates eleven-dimensional
supergravity and the two type-II supergravities in special
decompactification limits. But, as we stressed above, we are dealing
with a twelve-dimensional theory here, of which three coordinates are
compact, except that no field can depend on all of the three compact
coordinates. The theory has obviously two mass scales associated with
the KKA and KKB states. We return to them momentarily. Both S- and
T-duality are manifest, although the latter has become trivial as
the theory is not based on a specific IIA or IIB perspective. One has
the freedom to view the theory from a IIA or a IIB perspective and
interpret it accordingly. 

We should discuss the fate of the group ${\rm G}= {\rm SO}(1,1) \times
{\rm SL}(2,{\bf R})$ of pure supergravity after coupling the theory to
BPS multiplets. The central charges of the Kaluza-Klein states form a
discrete lattice, which is affected by this group. Hence, after
coupling to the BPS states, we only have a discrete subgroup that
leaves the charge lattice invariant. This is the group ${\rm
SL}(2,{\bf Z})$.  

The KKA and KKB states and their interactions can be
understood from the eleven-dimensional and IIB supergravity
perspective. Therefore we can deduce the following BPS mass formula,
\beq
M_{\rm BPS}(q_1,q_2,p) = m_{\scriptscriptstyle\rm KKA} \,{\rm
e}^{3\sigma/7} \,\vert q_\a \phi^\a \vert +  m_{\scriptscriptstyle\rm
KKB} \,{\rm e}^{-4\sigma/7} \,\vert p\vert\,,
\eeq
where $q_\a$ and $p$ refer to the integer-valued KKA and KKB charges,
respectively, and $m_{\scriptscriptstyle\rm KKA}$ and
$m_{\scriptscriptstyle\rm KKB}$ are two independent  mass scales. 
This formula can be compared to the membrane BPS formula
\eqn{BPS-membrane} in the eleven-dimensional frame. One then finds
that 
\beq
m^2_{\scriptscriptstyle \rm KKA}\, m^{~}_{\scriptscriptstyle\rm KKB}
\propto  T_{\rm m}\,,
\eeq
without field-dependent factors. 

The above example of a BPS-extended supergravity theory shows that one
obtains a dichotomic theory which can be regarded as a
twelve-dimensional field theory. Various decompactification limits
correspond to eleven-dimensional supergravity or ten-dimensional IIA/B 
supergravity. 
There are interesting questions regarding the field-theoretic coupling
of the fields associated with the BPS states. In the case at hand
these can in principle be answered, because the couplings can be deduced
from the coupling of the massive Kaluza-Klein fields in the
compactifications of eleven-dimensional and IIB supergravity.  
One such questions concerns the role of the local symmetry 
${\rm H}={\rm SO}(2)$ that one uses in the description of the 
SL(2)/SO(2) coset space for the nonlinear sigma model. There is a
composite gauge field associated with the group SO(2), which does not
correspond to additional  
degrees of freedom. To study this aspect in a little more detail,
let us consider a simplified example (worked out in collaboration with
I. Herger) illustrating the action of the 
hidden symmetries when the massive Kaluza-Klein modes are retained,
namely a nonlinear sigma model based on the coset space ${\rm
SL}(n,{\bf R})/{\rm SO}(n)$ in flat space-time. 
In the following, we will split the higher-dimensional coordinates
as $z^M = (x^\mu,y^m)$, where the $y^m$ parametrize a torus. The
degrees of freedom are thus contained  
in a matrix $\VC(x,y)\in {\rm SL}(n,{\bf R})$ transforming as
\beq
\VC(x,y) \rightarrow g\, \VC(x,y)\, h^{-1}(x,y)\,,
\eeq
where $g$ denotes a (constant) element of SL$(n,{\bf R})$ and $h(x,y)$
is a local SO$(n)$ transformation. In view of the gauge invariance that
depends on both $x$ and $y$, and the fact that we are dealing with a
group element,  the split into massive and
massless degrees of freedom is not entirely straightforward. 

The best approach is to write 
$\VC(x,y)$ as the product of two SL$(n,{\bf R})$ elements, 
\beq
\VC(x,y) = \VC_0(x)\,  \VC_1 (x,y)\,, \label{Vsplit}
\eeq
and to require that $\VC_0$ describes the massless modes in the torus
compactification. To do this, one can first fix the SO$(n)$ gauge freedom
and define a coset representative. Subsequently one considers the
logarithm of $\VC(x,y)$ and expands it in terms of Fourier modes on
the torus. Dropping the $y$-dependent modes in this expansion 
yields $\VC_0(x)$. However, $\VC_0(x)$ is itself a coset representative
so that it is defined up to multiplication by an $x$-dependent
SO$(n)$ transformation acting from the right. This leads to a
corresponding ambiguity for $\VC_1(x,y)$. Hence   
$\VC_0(x)$ parametrizes a nonlinear sigma model in the
lower-dimensional space, so that it transforms according to 
\beq
\VC_0(x) \rightarrow g\, \VC_0(x)\, h_0^{-1}(x)\,,
\eeq
where $h_0(x)$ is an $x$-dependent SO$(n)$ transformation, and 
$\VC_1(x,y)$ transforms under an $x$-dependent SO$(n)$ transformation 
from the left and, provided one again relaxes the original gauge condition,
under an $x$- and $y$-dependent SO$(n)$ transformation
from the right,
\beq
\VC_1(x,y) \rightarrow h_0(x) \,\VC_1(x,y) \,h_1^{-1}(x,y)\,
h_0^{-1}(x)\,,
\eeq
where we defined $h(x,y) = h_0(x)\, h_1(x,y)$. 
All the massive Kaluza-Klein degrees of freedom thus reside in
$\VC_1(x,y)$. 
The SO$(n)$ symmetry corresponding to 
$h_1(x,y)$ can now be fixed by going to a ``unitary gauge'',
\beq
\VC_1(x,y) = \exp \phi(x,y)\,,
\eeq
where $\phi(x,y)$ is a symmetric traceless $n\times n$ matrix, such
that 
$\VC_1(x,y)$ transforms under the residual $x$-dependent SO$(n)$
transformations according to 
\beq
\VC_1 (x,y) \rightarrow h_0(x)\, \VC_1 (x,y)\, h_0^{-1}(x)\,,\quad
\phi(x,y) \rightarrow h_0(x)\, \phi (x,y)\, h_0^{-1}(x)\,.
\eeq
Thus the massive fields $\phi(x,y)$ transform covariantly under
$x$-dependent SO$(n)$ gauge transformations but not under ${\rm
SL}(n,{\bf R})$.  

The split \eqn{Vsplit} of $\VC(x,y)$ exhibits clearly how the 
massive Kaluza-Klein degrees of freedom behave with respect to 
the local symmetries of the massless theory. 
To describe the Lagrangian we consider the SL$(n,{\bf R})$ Lie-algebra 
valued expression
\bea
P_M + Q_M &:=& \VC^{-1} \partial_M \VC  \nn
      &=& \VC_1^{-1} P^0_M \VC_1 +  \VC_1^{-1} D^0_M \VC_1 + Q^0_M \,, 
\eea
where $Q_M$ and $P_M$ belong to the Lie algebra of SO$(n)$ and its 
complement, respectively, in the Lie algebra of ${\rm SL}(n,{\bf R})$. 
Splitting the index $M$ into $\mu$ and $m$ as before, we have 
the $y$-independent quantities
\beq
P^0_\mu + Q^0_\mu := \VC_0^{-1} \partial_\mu \VC_0 
\eeq
(obviously, $Q^0_m = P^0_m =0$). The derivative $D^0_M$ is covariant 
with respect to $x$-dependent ${\rm SO}(n)$ gauge transformations,
\beq
D_\mu^0 \VC_1 := \partial_\mu \VC_1 + [ Q^0_\mu, \VC_1]\,,\qquad D_m^0
\VC_1 := \partial_m \VC_1  \,. 
\eeq
To write down an action coupling the massless sector and the massive
Kaluza-Klein modes in an SO$(n)$ invariant way, we expand
\beq
P_M = P_M^0 + D_M^0 \phi + [P_M^0,\phi] + \ft12 [[P_M^0,\phi],\phi] +
\ft12 [D_M^0 \phi, \phi] + \cdots\,, 
\eeq
projected on the complement of the Lie algebra of SO$(n)$. Because the
${\rm SL}(n,{\bf R})/{\rm SO}(n)$ coset space is symmetric some of the
terms in $P_M$ will trivially vanish. What remains is to substitute
the expression for $P_M$ into 
\beq
L = -\bruch{1}{2} {\rm Tr} \, (P_\mu^2)
    -\bruch{1}{2} {\rm Tr} \, (P_m^2)\,,
\eeq
which will lead to an action that is non-polynomial in $\phi$. Let
us repeat, however, that this action is invariant under  
$x$-dependent gauge transformations, as well as under a global
SL$(n,{\bf R})$ symmetry which acts exclusively in the massless sector. 
Once we fix an SO$(n)$ gauge, the SL$(n,{\bf R})$ symmetry becomes
nonlinearly realized and acts also on the massive fields. 

Before fixing an SO$(n)$ gauge, the SL$(n,{\bf R})$ symmetry does not
act on the massive modes in this simplified model. This is not so when
the SL$(n,{\bf R})$ originates from the dimensional reduction in  
the more complicated models based on (super)gravity in higher
dimensions.
Upon performing a Kaluza-Klein reduction (not a truncation!) on the
torus $T^n$, the global  
symmetry will still act on the massive modes, but it will be 
broken to an arithmetic subgroup such as G$(\bf{Z})$. To see how 
this comes about, recall that ${\rm G}={\rm SL}(n,{\bf R})$ and
${\rm H}={\rm SO}(n)$ are precisely the symmetries that one obtains upon
dimensional reduction of pure Einstein theory on a torus $T^n$.
As we saw in the foregoing section, the Kaluza-Klein gauge field 
${B_\mu}^m$ couples via the derivative operator (\ref{Dmu}), 
$$
\DC_\mu = \partial_\mu - {B_\mu}^m \partial_m
$$
(with vanishing affine connection for the torus).
When the theory is compactified on a torus $T^n$, the derivative
operator $\partial_m$ will only admit discrete eigenvalues
${\bf q} = (q_1,\ldots,q_m)$. These eigenvalues lie on an $n$-dimensional
lattice, the lattice of Kaluza-Klein charges. It is the presence of
this lattice that leads to the breaking SL$(n,{\bf R})\rightarrow {\rm
SL}(n,{\bf Z})$: the  
group SL$(n,{\bf Z})$ acts on the vectors $\bf q$ labeling the 
Kaluza-Klein modes, rather than on the fields themselves. The massless
modes have ${\bf q}=0$ and transform under SL$(n,{\bf Z})$ in the way
described above for the non-linear sigma model. 


Further results along these lines will be published elsewhere.

\vspace{2mm}
\noindent
{\bf Acknowledgments:} The Kaluza-Klein reduction of nonlinear
sigma models was obtained in collaboration with Ivan Herger. We thank 
the organizers of the G\"ursey Memorial Conference for inviting us to
a pleasant and stimulating meeting. This work is supported in part by the
European Commission RTN programmes HPRN-CT-2000-00122 and 
HPRN-CT-2000-00131 (in the latter H.N. is associated with U. Bonn).


\end{document}